\documentclass[aps,prb,twocolumn,10pt,floatfix]{revtex4-1}
\usepackage{ucs}
\usepackage[utf8x]{inputenc}  
\usepackage[T1]{fontenc}    
\usepackage[english]{babel}
\usepackage{graphicx}
\usepackage{hyperref}

\begin{document}
\title{Stray-field imaging of magnetic vortices with a single diamond spin}
\author{L.~Rondin$^{1,\dag}$}
\author{J.-P.~Tetienne$^{1,2,\dag}$}
\author{S.~Rohart$^{3}$}
\author{A.~Thiaville$^{3}$}
\author{T. Hingant$^{2}$}
\author{P.~Spinicelli$^{1}$}
\author{J.-F.~Roch$^{2}$}
\author{V.~Jacques$^{1,2}$}
\email{vjacques@ens-cachan.fr}
\affiliation{$^{1}$Laboratoire de Photonique Quantique et Mol\'eculaire, Ecole Normale Sup\'erieure de Cachan and CNRS UMR 8537, 94235 Cachan Cedex, France}
\affiliation{$^{2}$Laboratoire Aim\'e Cotton, CNRS, Universit\'e Paris-Sud and ENS Cachan, 91405 Orsay, France.}
\affiliation{$^{3}$Laboratoire de Physique des Solides, Universit\'e Paris-Sud and CNRS UMR 8502, 91405 Orsay, France}
\altaffiliation{These authors contributed equally to this work.}

\begin{abstract}
Despite decades of advances in magnetic imaging, obtaining direct, quantitative information with nanometer scale spatial resolution remains an outstanding challenge. Recently, a new technique has emerged that employs a single nitrogen-vacancy (NV) defect in diamond as an atomic-size magnetometer. Although NV magnetometry promises significant advances in magnetic imaging, the effectiveness of the technique, when applied to realistic magnetic nanostructures, remains to be demonstrated. Here we use a scanning NV magnetometer to image a magnetic vortex, which is one of the most iconic object of nanomagnetism, owing to the small size ($\sim 10$~nm) of the vortex core. We report three-dimensional, vectorial, and quantitative measurements of the stray magnetic field emitted by a vortex in a ferromagnetic square dot, including the detection of the vortex core. We find excellent agreement with micromagnetic simulations, both for regular vortex structures and for higher-order magnetization states. These experiments establish scanning NV magnetometry as a practical and unique tool for fundamental studies in nanomagnetism.
\end{abstract}

\maketitle

\noindent {\bf Introduction.} Although a remarkable number of magnetic microscopy techniques have been developed over the last decades, imaging magnetism at the nanoscale remains a challenging task since it requires a combination of high spatial resolution and sensitivity~\cite{Freeman_Science2001}. A first approach consists in directly mapping the sample magnetization, which implies sending and collecting back test particles whose interaction with matter has a magnetization-dependent term. This approach provides the highest spatial resolution to date, down to the atomic-scale for spin-polarized scanning tunneling microscopy~\cite{Wachowiak2002} and about $10$~nm in transmission X-ray microscopy~\cite{Fischer2012}. However, these techniques require highly complex experimental apparatus and a dedicated sample preparation so that the particles can reach and escape from the tested region without perturbations. To observe magnetic samples in their real, unprepared state, a more suited approach consists in mapping the magnetic stray field generated outside the sample, even if this method cannot uniquely determine the actual sample magnetization~\cite{IEEE}. Furthermore, the spatial resolution is then limited both by the probe size and its distance to the sample. Among many stray field microscopy techniques~\cite{Kirtley2010}, magnetic force microscopy (MFM) has become ubiquitous, as it provides a spatial resolution below 50 nm~\cite{Martin1987} and operates under ambient conditions without any specific sample preparation. It was for instance the first method that allowed the observation of the core of a magnetic vortex in a thin ferromagnetic disk~\cite{Shinjo2000}. However, due to the intrinsic magnetic nature of the probe, MFM is known to be perturbative and not easily quantitative~\cite{Garcia2001}, therefore limiting its field of applications. \\ 
\indent Recently, a magnetometer based on the magnetic response of a single nitrogen-vacancy (NV) defect in diamond has been proposed~\cite{Taylor2008,Degen2008,Balasubramanian2008,Maze2008}, which promises significant advances in magnetic imaging. Indeed, it provides non-perturbing and quantitative measurements of the stray magnetic field, with an unprecedented access to low fields combined with an atomic-sized detection volume~\cite{Rondin2012,Maletinsky2012}. In this letter, we use NV-based magnetometry to measure the stray field emanating from magnetic vortices in nanostructured ferromagnetic thin films. Such structures, which are of great interest both for fundamental studies in nanomagnetism~\cite{Antos2008} and for applications such as non-volatile magnetic storage~\cite{Drews2009,Pigeau2010} and microwave generation~\cite{Pribiag2007}, have been investigated using a wide range of microscopy techniques~\cite{Shinjo2000,Raabe2000,Chung2010,Wachowiak2002,Choe2004,Vansteenkiste2009}. However, obtaining quantitative information that is directly comparable to theory remains a challenging task. Further, imaging the vortex core, which can be as small as $10$~nm and plays a crucial role in the vortex dynamics, is a long-standing goal that has been reached by very few methods only~\cite{Shinjo2000,Wachowiak2002,Fischer2012}. Here we show that scanning NV magnetometry enables to quantitatively map the stray field above a thin ferromagnetic square in a vortex state, revealing in 3D the full structure of the magnetic field distribution, including the detection of the vortex core. Furthermore, we demonstrate that the vectorial and quantitative nature of the measurement provide direct comparisons with micromagnetic simulations. This work thus opens new avenues for fundamental studies in nanomagnetism and spintronics.

\begin{figure}
\begin{center}
    \includegraphics[width=0.49\textwidth]{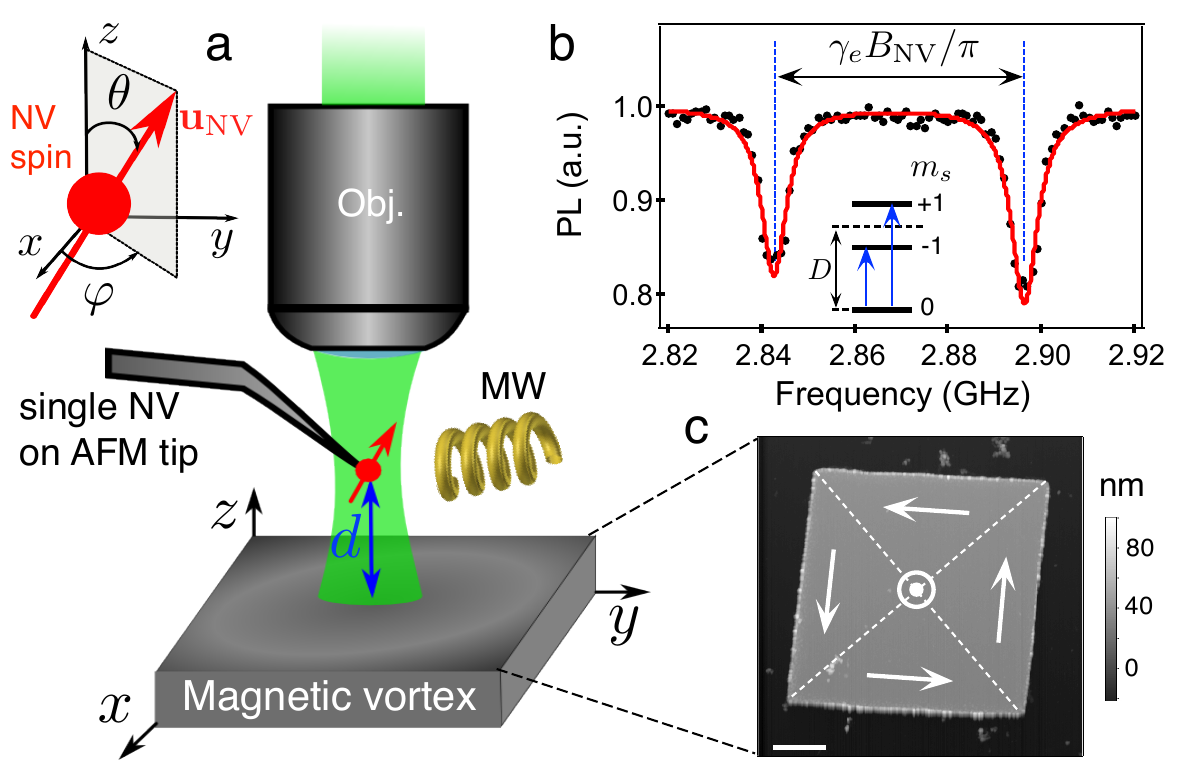}
\end{center}
\caption{{\bf Principle of the experiment.} ({\bf a})-The apex of an AFM tip is functionalized with a 20-nm diamond nanocrystal hosting a single NV defect. A microscope objective placed on top of the AFM tip is used both for exciting and collecting the NV defect spin-dependent PL. A microwave (MW) field is generated by an antenna approached in the vicinity of the NV defect. The NV defect quantization axis $\mathbf{u}_{\rm NV}$, which gives the measured magnetic field component, is described by the angles $\theta$ and $\varphi$ in a spherical coordinate system. ({\bf b})-ESR spectrum of the magnetic probe recorded by monitoring the NV defect PL intensity while sweeping the frequency of the MW field. The ESR splitting is proportional to the projection of the magnetic field along the NV axis, here $B_{\rm NV}=0.9$~mT. ({\bf c})-Typical AFM image of a thin square of Fe$_{20}$Ni$_{80}$. The white arrows indicate the curling magnetization of the vortex state. Scale bar, $1 \ \mu$m.}
\label{Fig1}
\end{figure}

\noindent {\bf Principle of scanning NV magnetometry.} The scanning NV magnetometer combines an optical confocal microscope and an atomic force microscope (AFM), all operating under ambient conditions (see Methods). As sketched in Figure~\ref{Fig1}(a), a diamond nanocrystal hosting a single NV defect is grafted at the apex of the AFM tip and used as an atomic-sized magnetic sensor~\cite{Rondin2012}. The NV defect ground state is a spin triplet with a zero-field splitting $D=2.87$ GHz between a singlet state $m_{s}=0$ and a doublet $m_{s}=\pm 1$, where $m_s$ denotes the spin projection along the intrinsic quantization axis of the NV defect $\mathbf{u}_{\rm NV}$ (Fig.~\ref{Fig1}(a)). Under optical illumination, the NV defect is efficiently polarized into the $m_{s}=0$ spin sublevel and exhibits a spin-dependent photoluminescence (PL)~\cite{Manson_PRB2006}. These combined properties enable the detection of electron spin resonance (ESR) by optical means~\cite{Gruber_Science1997}. A typical ESR spectrum of the NV sensor placed in a static magnetic field is shown in Figure~\ref{Fig1}(b). In the limit of weak magnetic fields ($<3$~mT), the spin quantization axis remains fixed by the NV defect axis, and the two ESR frequencies are given by $\nu_R =D\pm\gamma_e B_\mathrm{NV}/ 2\pi$, where $ B_\mathrm{NV}$ is the magnetic field projection along the NV axis and $\gamma_e$ is the electron gyromagnetic ratio (Fig.~\ref{Fig1}(b)). Note that strain-induced splitting of the $m_{s}=\pm 1$ spin sublevels has been omitted for clarity purpose (see Supplementary Information). Measurement of the magnetic field through Zeeman shifts of the ESR frequency is therefore quantitative and non perturbing, since the dipolar field from a single NV defect is as low as $1 \ \mu$T at $10$~nm distance~\cite{Taylor2008}.\\

\noindent {\bf Stray field imaging of magnetic vortices in ferromagnetic films.} In the following, we use the scanning NV magnetometer to map the magnetic stray field generated by magnetic vortices in ferromagnetic thin films. More precisely, we investigate squares of Fe$_{20}$Ni$_{80}$ with a thickness of $50$ nm and a $5\ \mu$m side length. This magnetic structure is characterized by a curling in-plane magnetization with a vortex core in the center where the magnetization points out of the plane (Fig.~\ref{Fig1}(c)). Such a magnetic vortex is an ideal object for evaluating the sensitivity and spatial resolution of any advanced magnetic microscopy technique since the size of the vortex core can reach $10$~nm~[\onlinecite{Fischer2012}]. Furthermore this magnetic structure is topologically stable and therefore appears as an interesting candidate for memory cells in non-volatile data-storage devices~\cite{Drews2009,Pigeau2010}.\\
\begin{figure}[b]
\begin{center}
    \includegraphics[width=0.49\textwidth]{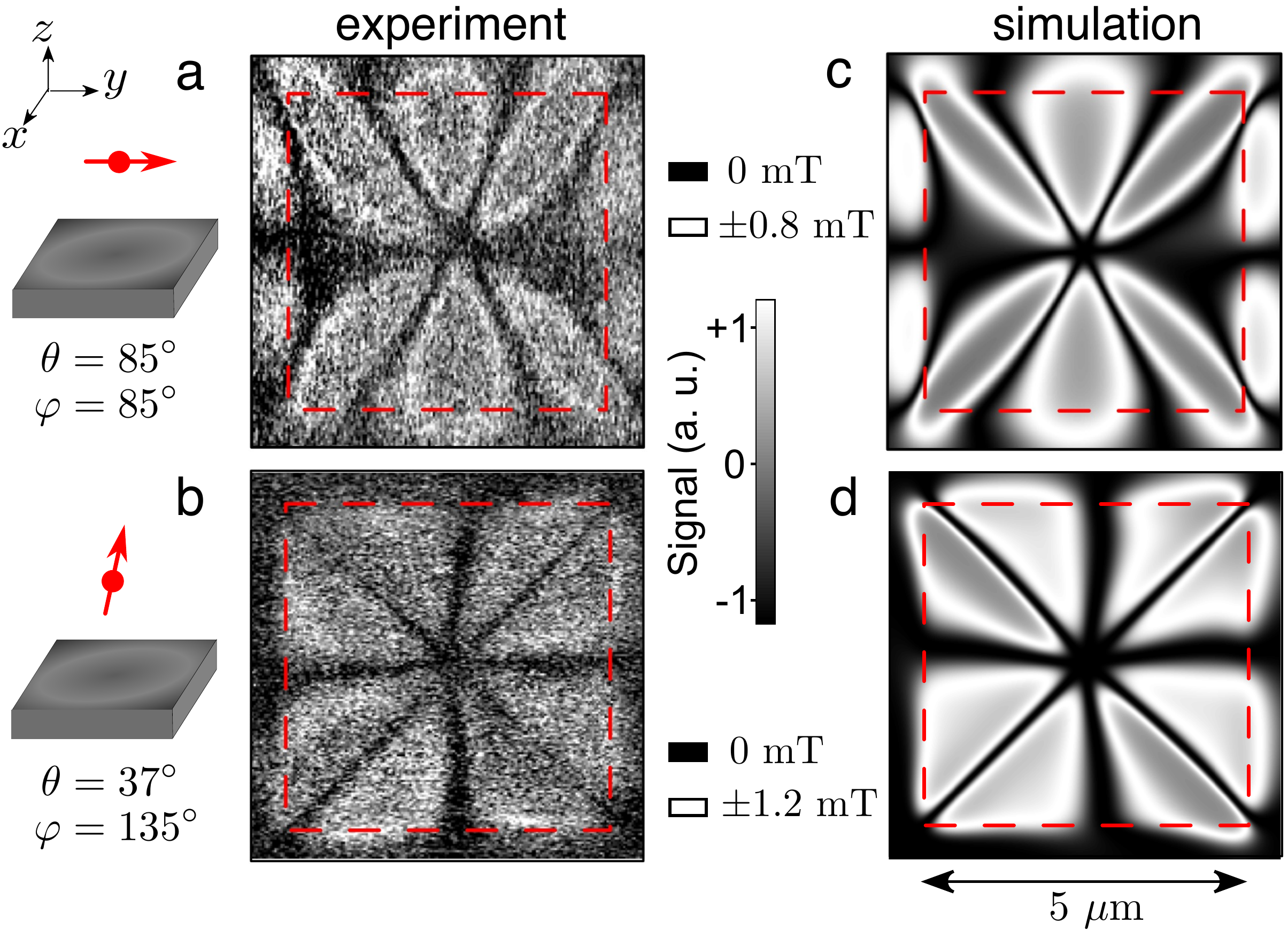}
\end{center}
\caption{{\bf Stray field imaging of a square of Fe$_{20}$Ni$_{80}$}. ({\bf a}),({\bf b})-Dual-iso-B images recorded with a probe-to-sample distance $d=400$~nm in (a) and $d=300$~nm in (b). The red squares depict the $5 \ \mu$m magnetic structure. In (a), the NV is oriented mainly along the $y$-axis ($\theta=85^{\circ}$, $\varphi=85^{\circ}$) and in (b) the NV defect has a strong component along the $z$-axis ($\theta=37^{\circ}$, $\varphi=135^{\circ}$). Images correspond to $200\times 200$~pixels, with a 30-nm pixel size and a $60$~ms acquisition time per pixel. ({\bf c}),({\bf d})-Simulated magnetic field maps corresponding to (a) and (b), respectively.}
\label{Fig2}
\end{figure}
\indent The NV probe is first approached at a distance $d\approx$~300 nm above the magnetic vortex structure. Magnetic field mapping is performed by scanning the sample while measuring the difference of NV defect PL intensity for two fixed microwave (MW) frequencies $\nu_1$ and $\nu_2$, applied consecutively at each point of the scan~\cite{Rondin2012}. This signal is positive if $\nu_R=\nu_1$, {\it i.e.} when the local field experienced by the NV defect is $B_{\rm NV,1}=\pm 2\pi(\nu_1-D)/\gamma_e$, and negative if $\nu_R=\nu_2$, {\it i.e.} for $B_{\rm NV,2}=\pm 2\pi (\nu_2-D)/\gamma_e$ (see Supplementary Information). The resulting image thus exhibits positive and negative signal regions corresponding to iso-magnetic field contours ($B_{\rm NV,1}$,$B_{\rm NV,2}$), as well as zero-signal regions for any other field projections. An example of such a dual-iso-B  image recorded above a ferromagnetic square is presented in Figure~\ref{Fig2}(a), revealing a flower-shaped magnetic field distribution. Independent measurement of the NV defect orientation with a calibrated magnetic field indicates that the NV defect is oriented mainly along the $y$ axis in this experiment (see Supplementary Information). Hence the magnetic map shown in Figure~\ref{Fig2}(a) corresponds to the field component that is parallel to the sample surface and along a side of the ferromagnetic square. By properly selecting the orientation of the single NV defect grafted at the apex of the AFM tip, any field component can be similarly measured. For instance, Figure~\ref{Fig2}(b) is essentially a map of the $z$-component, {\it i.e.} the out-of-plane magnetic field. These experiments establish NV-based magnetometry as a unique instrument providing quantitative and vectorial magnetic field images at the nanoscale.\\
\indent The measured stray field arises from N\'eel domain walls at the square diagonals which induce volume charges with opposite sign on each side of the wall. The general lobe structure of the magnetic distribution can therefore be qualitatively understood by considering the magnetic field created by an assembly of magnetic dipoles placed along the diagonals. To get more precise predictions, the sample magnetization distribution was first calculated through micromagnetic simulations using OOMMF software~\cite{oommf}, with a cell size of $5\times5\times5$~nm$^3$. Once the equilibrium magnetization state found, the magnetic field generated by the structure is computed by summing the contribution of all magnetization cells. This magnetic field is then projected along the experimentally measured NV axis, and the NV defect ESR response is finally applied to get the simulated dual-iso-B image (see Supplementary Information). As shown in Figure~\ref{Fig2}(c)\&(d), the simulations reproduce well the experimental data, including apparent asymmetries which stem from the imperfect alignment of the corresponding NV defects with respect to the $y$ and $z$ axis. By varying the probe-to-sample distance $d$, it is even possible, as shown in the Supplementary Information, to obtain an overview of the stray field in the half-space above the sample, thus providing a 3D mapping of the magnetic field distribution. \\

\begin{figure}
\begin{center}
    \includegraphics[width=0.49\textwidth]{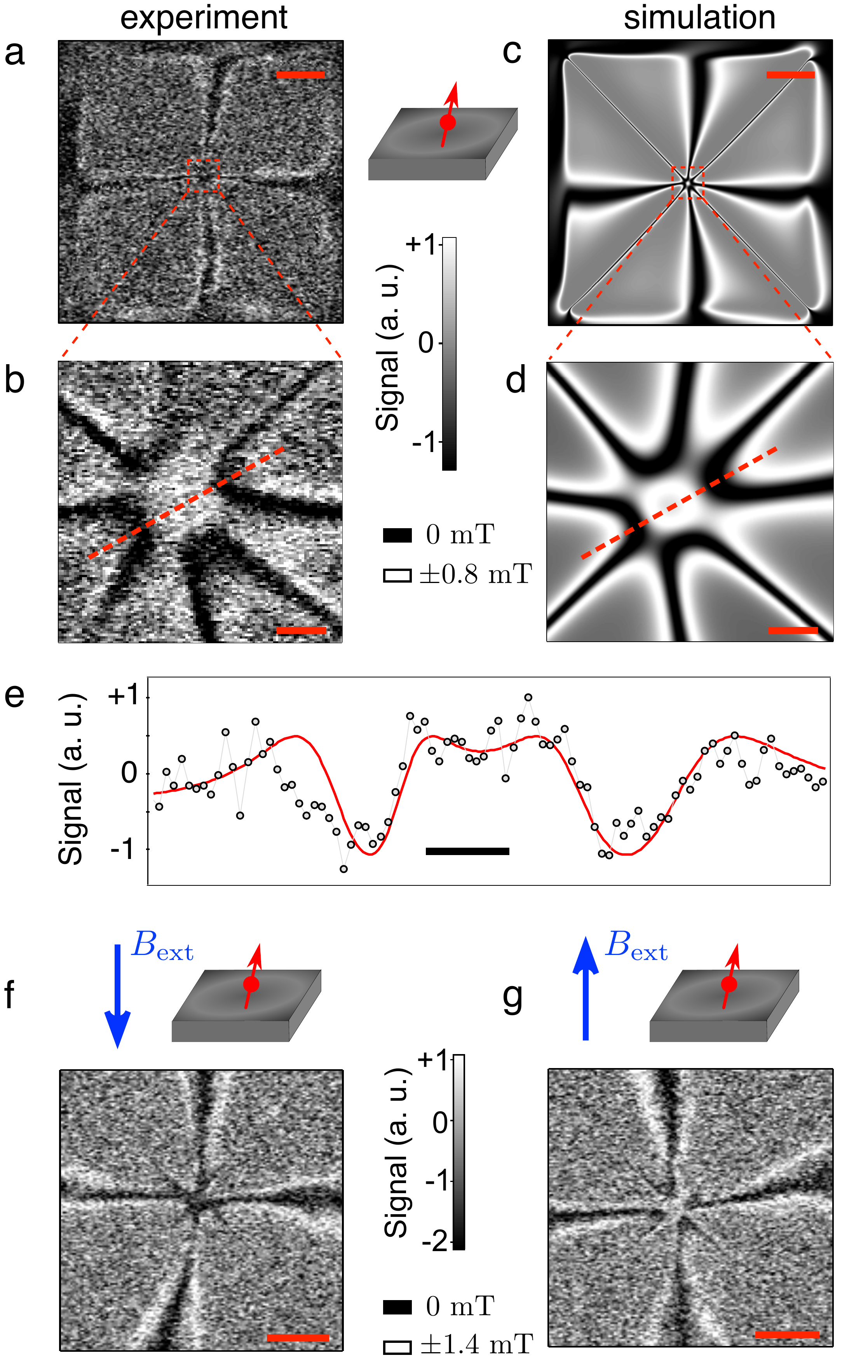}
\end{center}
\caption{{\bf Observation of the vortex core.}  ({\bf a})-Dual-iso-B image of the whole square obtained with the AFM operating in tapping mode. The same NV probe as in Fig.~\ref{Fig2}(b) is used ($\theta=37^{\circ}$, $\varphi=135^{\circ}$). Scale bar, $1 \ \mu$m. ({\bf b})-Zoom into the center of the structure enabling the detection of the stray field emanating from the vortex core magnetization. The image corresponds to $110\times 110$~pixels, with a $4.7$~nm pixel size and a $60$~ms acquisition time per pixel. Scale bar, $100$~nm. ({\bf c}),({\bf d})-Simulated magnetic field maps corresponding to (a) and (b), respectively, with a probe-to-sample distance $d=100$~nm. (e)-Line-cut taken along the red dashed line in (b) (markers) and (d) (solid line). Scale bar, $50$~nm. ({\bf f}),({\bf g})-Dual-iso-B image of the square center while applying a bias field with a projection along the NV axis $B_\mathrm{ext}=-0.8$ mT in (f) and $B_\mathrm{ext}=+0.8$ mT in (g). Scale bar, $500$~nm.}
\label{Fig3}
\end{figure}

\noindent {\bf Imaging the vortex core.} The magnetic images shown in Fig.~\ref{Fig2}(a)\&(b) are recorded with a probe-to-sample distance of several hundreds of nanometers. At such distances, the stray field linked to the out-of-plane vortex core magnetization cannot be detected. The vortex center therefore appears dark (zero field) both in the images and in the simulations. With the aim of observing the vortex core, the NV probe is brought as close as possible to the sample with the AFM operating in tapping mode. Figure~\ref{Fig3}(a) shows the dual-iso-B image of the whole square obtained with the NV defect probe used in Fig.~\ref{Fig2}(b), {\it i.e.} mainly oriented along the $z$ axis. By zooming in the center of the structure, the vortex core is revealed with an apparent size $L_{\rm core}\approx 100$~nm (Fig.~\ref{Fig3}(b)). Although the spatial resolution of scanning NV magnetometry is ultimately given by the atomic-sized detection volume, the effective resolving power is rather limited by the probe-to-sample distance, a common feature of any stray field microscopy technique. Indeed, numerical calculations indicate that the size of the out-of-plane vortex core magnetization is around $20$ nm at the sample surface. The probe-to-sample distance can be estimated by comparing the experimental results to magnetic images simulated at different distances from the sample surface. As shown in Figures~\ref{Fig3}(c),(d)\&(e), a good agreement is obtained for a distance $d\approx 100$~nm. We attribute this relatively large value to an imperfect positioning of the diamond nanocrystal at the apex of the AFM tip. A more precise control of the NV magnetic sensor position could be achieved by using diamond nanopillar-probes~\cite{Maletinsky2012}. Although the simulation fairly agrees with experimental data, we note that the sharp structures lying along the square diagonals (Fig.~\ref{Fig3}(c)) are not observed in the experimental images (Fig.~\ref{Fig3}(a)). In those regions, the stray field generated by N\'eel domain walls is strong enough to induce a mixing of the NV defect electron spin sublevels which results in an overall reduction of ESR contrast~\cite{Tetienne2012} (see Supplementary Information). \\

\noindent {\bf Chirality and polarity of the vortex state.} The vortex structure is commonly characterized by two independent binary properties. The chirality $c$ determines  whether the in-plane magnetization is curling clockwise ($c=+1$) or counterclockwise ($c=-1$), while the polarity $p$ of the vortex indicates the upward ($p=+1$) or downward ($p=-1$) magnetization of the core. The apparent pairing of the zero-field contours around the vortex core observed in Figure \ref{Fig3}(b) directly enables us to determine that $c\times p=+1$ (see Supplementary Information). We note that only an absolute value of the field is measured, preventing a direct measurement of the core polarity without additional measurements. This limitation could be overcome by addressing selectively one of the two ESR transitions through circularly polarized MW excitation~\cite{Alegre2007}. Here, we gain further insights into the vortex structure by applying an external bias magnetic field along the $z$-axis with a projection $B_{\rm ext}$ along the NV axis. The amplitude of this external magnetic field is chosen weak enough to avoid modifying the sample magnetization, so that $B_\mathrm{ext}$ just adds up to the regular stray field of the vortex structure. Figures \ref{Fig3}(f)\&(g) show magnetic images obtained with $B_{\rm ext}=-0.8$ mT and $B_{\rm ext}=+0.8$ mT, respectively. The stray field of the vortex core turns dark -- the net magnetic field vanishes -- with the negative bias field, which reveals that the vortex core magnetization points upward $p=+1$, and indicates a stray field emanating from the core $B_{\rm core}\approx+0.8$ mT. Conversely, the core field is increased with a positive bias field since it adds up to $B_{\rm core}$. From this polarity measurement, the chirality $c=+1$ can finally be deduced leading to a full characterization of the vortex structure.\\

\begin{figure}
\begin{center}
    \includegraphics[width=0.49\textwidth]{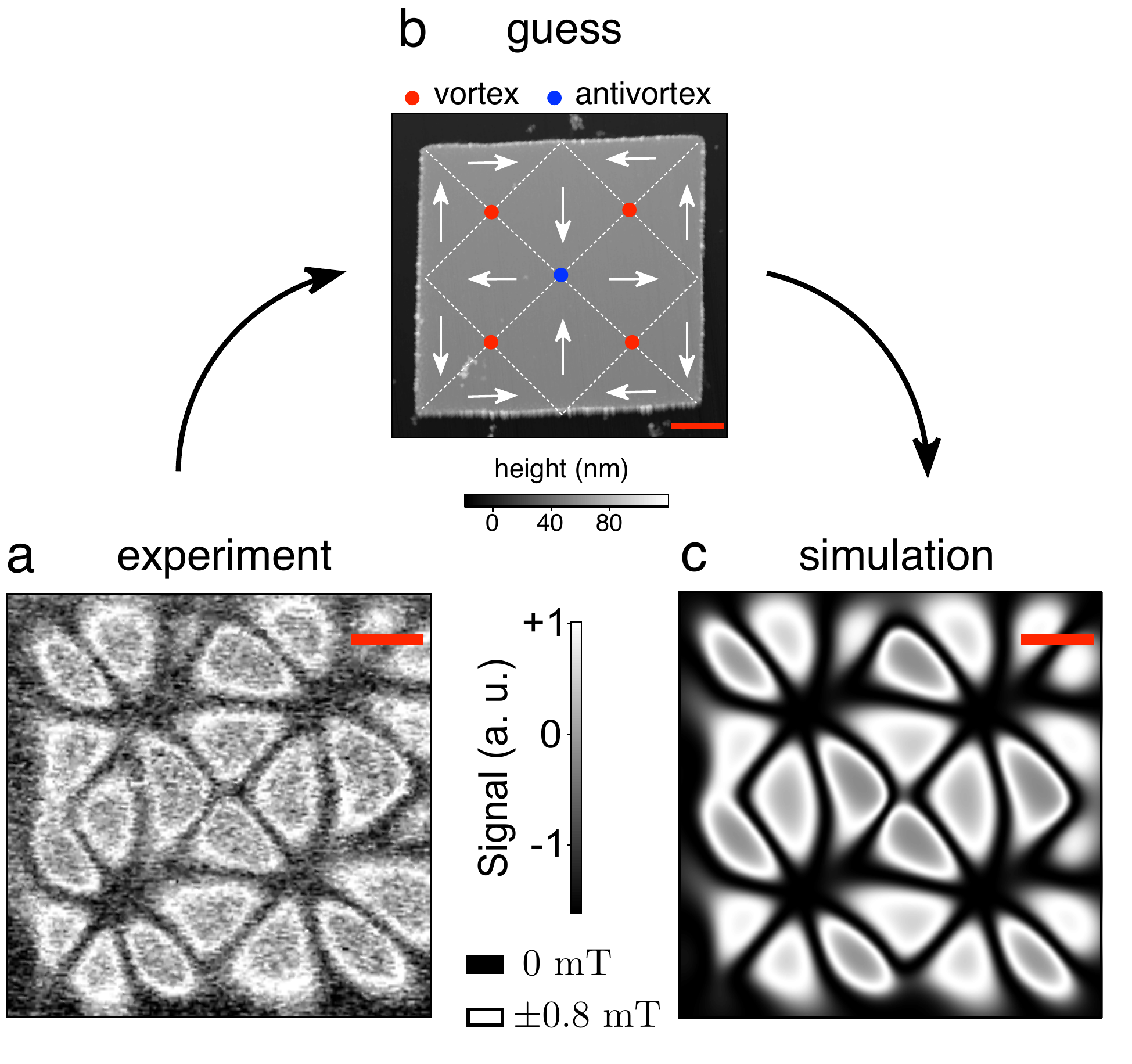}
\end{center}
\caption{{\bf Higher-order magnetization distribution.} ({\bf a})-Magnetic field distribution recorded above a square of Fe$_{20}$Ni$_{80}$ for a probe-to-sample distance $d=300$~nm. The NV defect orientation is given by $\theta=65^{\circ}$ and $\varphi=30^{\circ}$. ({\bf b})-AFM image of the square of Fe$_{20}$Ni$_{80}$. The white arrows depict the higher-energy magnetization structure, with four magnetic vortices (red dots) distributed around one anti-vortex at the center (blue dot). ({\bf c})-Simulated image considering a magnetization following the structure shown in (b). Scale bar, $1 \ \mu$m.}
\label{Fig4}
\end{figure}

\noindent {\bf Imaging higher-order magnetization distribution.}  As a final experiment, we study a higher-order magnetization distribution on a Fe$_{20}$Ni$_{80}$ square dot. As shown in Figure~\ref{Fig4}(a), highly non trivial magnetic field distributions are observed above some square ferromagnetic structures of the sample. Obviously these particular squares are not in the regular vortex distribution studied previously. One way to determine the magnetization distribution is to make an assumption about it, compute the expected dual-iso-B-image, and compare to experimental data. Here the four near-symmetry points distributed around the square center suggest a magnetization state comprising four vortices distributed around one anti-vortex at the center (Fig.~\ref{Fig4}(b)), which is a known equilibrium state in such ferromagnetic structures~\cite{Shigeto2002}. Comparison between the simulated (Fig.~\ref{Fig4}(c)) and measured magnetic images unambiguously validates our hypothesis that this particular square exhibits a four-vortex-one-antivortex magnetization state~\cite{Note}. \\

\noindent {\bf Discussion.} These experiments illustrate that scanning NV magnetometry enables robust comparisons with micromagnetic simulations and therefore appears as a powerful tool for fundamental studies in nanomagnetism and, more generally, nanoscience. Indeed, the direct measurement of weak magnetic fields, without perturbation and with a nanoscale resolution, may allow clearing some important issues, such as the nature of magnetic domain walls in ultrathin films with perpendicular anisotropy, which controls their current-induced motion~\cite{Khvalkovskiy13}. Bloch- and N\'eel-type domain walls exhibit distinct stray field distributions which could be discriminated by scanning NV magnetometry. As an example, micromagnetic simulations indicate that the typical stray field $50$~nm above a Bloch wall in a $0.6$~nm Co film with perpendicular anisotropy is $\approx 5$~mT, that changes by $\pm 0.5$~mT for a N\'eel wall (see Supplementary Information). Such a difference is large enough to allow NV magnetometry to experimentally determine the nature of the domain wall and analyse the relevance of Dzyaloshinskii-Moriya interactions in ultrathin magnetic films in contact with non magnetic layers~\cite{Bode2007}, which is expected to stabilize domain wall into the N\'eel configuration~\cite{Thiaville2012}.\\
\indent The detection of exotic magnetic structures known as skyrmion lattices~\cite{Heinze2011} is another example of a challenge that may be met by scanning NV magnetometry. On a broader perspective, the observation of the stray fields created by persistent currents in micro- and nanostructures where the phase coherence of the carriers is preserved, a quantum realization of the naive Amperian currents proposed two centuries ago to explain the magnetism of matter, appears also within reach of scanning NV magnetometry.\\

The authors acknowledge C.~Dal Salvio and K.~Karrai for fruitful discussions. This work was supported by the Agence Nationale de la Recherche (ANR) through the project D{\sc iamag}, and by C'Nano \^Ile-de-France (contracts M{\sc agda} and N{\sc anomag})

\section*{Methods}

 {\small {\bf Experimental setup and sample fabrication.} The experimental setup combines a tuning-fork-based atomic force microscope (AFM) and a confocal optical microscope (attoAFM/CFM, Attocube Systems), all operating under ambient conditions. A detailed description of the setup as well as the method to graft a diamond nanocrystal at the apex of the AFM tip can be found in Ref.\cite{Rondin2012}. We use commercially available diamond nanocrystals (SYP 0.05, Van Moppes SA, Geneva), in which single NV defects were created through high energy electron irradiation ($13.6$~MeV) followed by thermal annealing at $800^{\circ}$C. The irradiated nanocrystals were finally oxidized in air at $550^{\circ}$C during two hours. This procedure enables to reduce the size of the nanodiamonds and leads to an efficient charge state conversion of the created NV defects into the negatively-charged state~\cite{Rondin2010}. For the experiments reported in the main article, the size of the nanodiamonds were in the $\sim 20$~nm range, as verified using AFM images before grafting the nanodiamond at the apex of the tip. The unicity of the NV defect was checked through measurements of antibunching in the second order correlation function $g^{(2)}(\tau)$ of the NV defect photoluminescence (PL), using a standard Hanbury Brown and Twiss interferometer.\\
\indent For electron spin resonance (ESR) spectroscopy, a microwave (MW) excitation is applied through a 20 $\mu$m copper wire directly spanned on the magnetic sample. The $5\times5$~$\mu$m$^2$ square dots were patterned on a silicon substrate using electron-beam lithography followed by evaporation of $50$~nm of Fe$_{20}$Ni$_{80}$ and lift-off. All experiments were performed at zero external magnetic field except the measurement of the vortex core polarity which required to apply a bias field. The acquisition time of all magnetic images was set to $60$~ms per pixel corresponding to a total acquisition time of $10$~mn for a $100\times 100$~pixels image.
}

\section*{Supplementary Information}

\subsection{ESR frequencies vs. magnetic field}
In this section, we give the analytical expression of the NV defect ESR frequencies as a function of the applied magnetic field, while taking into account strain effects of the diamond lattice. 

\begin{figure}[t]
\begin{center}
    \includegraphics[width=0.47\textwidth]{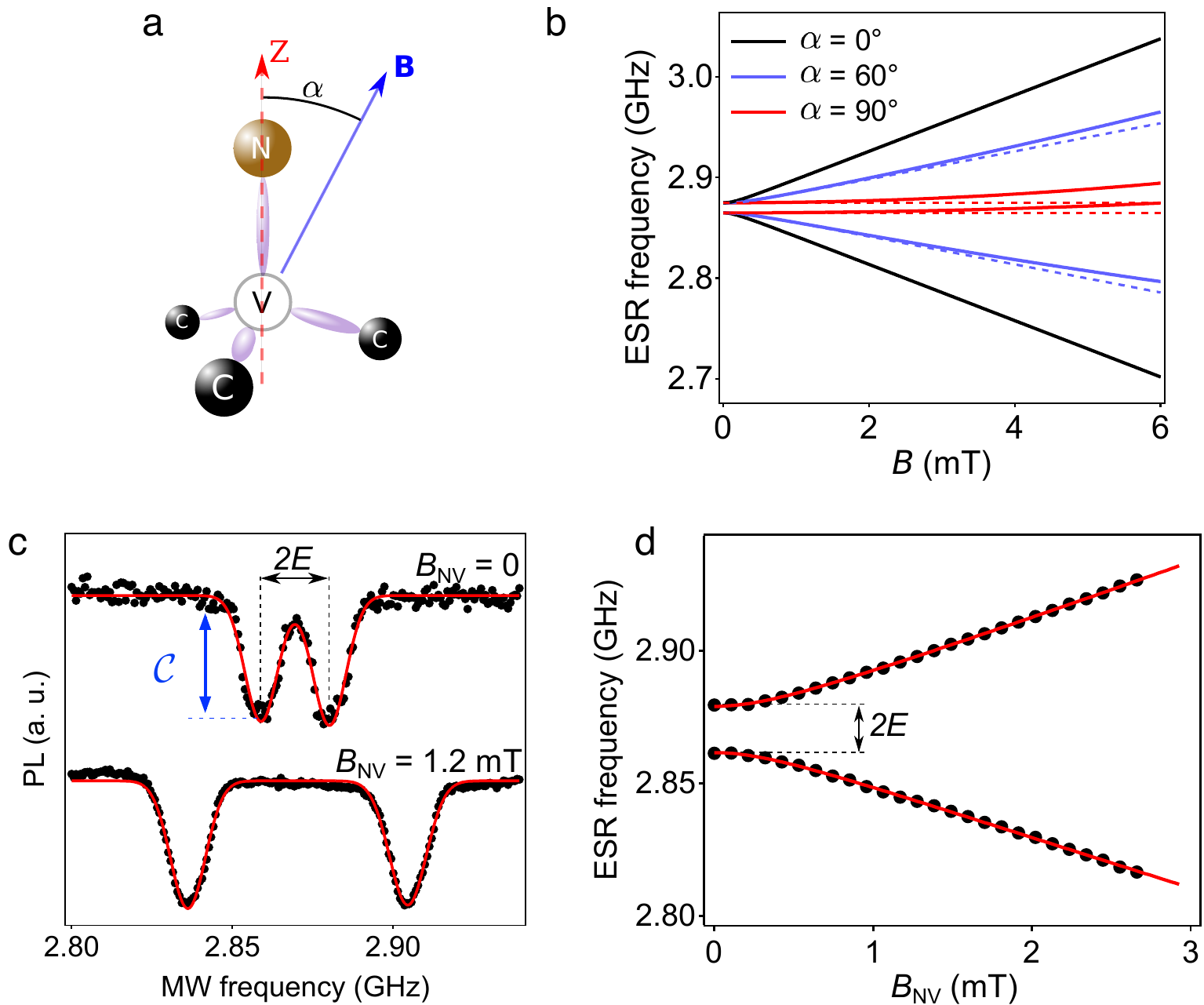}
\end{center}
\caption{(a)-Schematic of the NV defect. A magnetic field {\bf B} is applied with an angle $\alpha$ with respect to the NV defect axis. (b)-ESR frequencies $\nu_{R,\pm}$ as a function of the magnetic field amplitude $B=\Vert {\bf B}\Vert$ for different angles $\alpha$. The solid lines are obtained by diagonalizing the full Hamiltonian described by Equation~(\ref{eq:hamiltonian}), while using $D=2.87$~GHz and $E=5$~MHz. The dashed lines are obtained by using Equation~(\ref{eq:ESRfreq3}). (c)-Optically-detected ESR spectra of a single NV defect placed at the apex of the AFM tip for $B_\mathrm{NV}=0$ mT and $B_\mathrm{NV}=1.2$ mT (down-shifted for clarity). The solid lines are data fitting with Equation~(\ref{eq:ESR}) which allows us to determine the ESR linewidth $\Delta\nu=13$ MHz and the ESR contrast ${\cal C}=0.11$, for this particular NV defect. (d)-ESR frequencies as a function of $B_{\rm NV}$, with $\alpha=44^\circ$. The solid lines are the predictions of Equation (\ref{eq:ESRfreq3}) with the zero-field parameters $D=2869.5$ MHz and $E=10.5$ MHz.}
\label{fig:ESR_vs_B}
\end{figure}

The ground-state spin Hamiltonian of the NV defect electron spin ($S=1$) reads 
\begin{eqnarray} \label{eq:hamiltonian}
\mathcal{H} = h DS_Z^{2}+h E(S_X^{2}-S_Y^{2})+\hbar\gamma_{e}\mathbf{B}\cdot\mathbf{S}
\end{eqnarray}
where $h$ is the Planck constant, $D$ is the zero-field splitting, $E$ the strain splitting, $Z$ the NV defect axis, $\gamma_{e}$ the electron gyromagnetic ratio, and $\mathbf{B}$ the local magnetic field [Fig. \ref{fig:ESR_vs_B}(a)]. By numerically computing the eigenenergies of $\mathcal{H}$, the two ESR frequencies $\nu_{R,\pm}$ can be calculated for any magnetic field $\mathbf{B}$. For instance, the ESR frequencies $\nu_{R,\pm}$ are plotted in Figure~\ref{fig:ESR_vs_B}(b) as a function of the magnetic field amplitude $B=\Vert {\bf B}\Vert$ for various angles $\alpha$ between {\bf B} and the NV defect axis.   

With the aim of obtaining a simple analytical expression for $\nu_{R,\pm}$, the Hamiltonian $\mathcal{H}$ is written 
\begin{eqnarray} \label{eq:hamiltonian2}
\mathcal{H} = \mathcal{H}_\parallel + \mathcal{H}_\perp+\mathcal{H}_{\rm strain}
\end{eqnarray}
with
\begin{equation}
\left\{
\begin{array}{rl} \label{eq:hamiltonian3}
\mathcal{H}_\parallel & =  h DS_Z^{2}+\hbar\gamma_{e}B_{\rm NV} S_Z \\ 
\mathcal{H}_\perp & =  \hbar\gamma_{e}(B_X S_X+B_Y S_Y)\\
\mathcal{H}_{\rm strain} & =  h E(S_X^{2}-S_Y^{2}) \ ,
\end{array}
\right.
\end{equation}
where we introduced the notation $B_{\rm NV}=B_Z$ as used in the main article. Typical values for the zero-field parameters in diamond nanocrystals are $D\approx2.87$ GHz and $E\approx5$ MHz. One can therefore distinguish different regimes depending on the value of $B_\parallel=|B_{\rm NV}|$ and $B_{\perp}=\sqrt{B_X^2+B_Y^2}$, which are the parallel and transverse components of the magnetic field with respect to the NV axis, respectively. 

Focusing on weak magnetic field amplitudes, we consider that $\mathcal{H}_\perp\ll  \mathcal{H}_\parallel$. The ESR frequencies are then simply given by 
\begin{equation} \label{eq:ESRfreq3}
\nu_{R,\pm}(B_{\rm NV})=D\pm\sqrt{(\gamma_{e}B_{\rm NV}/2\pi)^2+E^{2}} \ .
\end{equation}
This formula can be used whenever the transverse component of the magnetic field $B_\perp$ is much smaller than $2\pi D/\gamma_e\approx 100$ mT. This is illustrated in Figure \ref{fig:ESR_vs_B}(b), where the prediction of Equation (\ref{eq:ESRfreq3}) is plotted in dotted lines together with the full calculation (solid lines). Obviously the approximate formula is relevant for field amplitudes in the range 0-2 mT, which is the range we address in this work. In addition, if $\gamma_e B_\parallel /2\pi \gg E$, then $\mathcal{H}_{\rm strain}$ can be neglected and the ESR frequencies become
\begin{equation} \label{eq:ESRfreq1}
\nu_{R,\pm}(B_{\rm NV})=D\pm\gamma_{e}B_{\rm NV}/2\pi \ ,
\end{equation}
as indicated in the main paper.

Experimentally, the ESR frequencies are inferred from the optically-detected ESR spectrum, i.e. the NV defect PL rate ${\cal R}(\nu,B_{\rm NV})$ as a function of the MW frequency $\nu$. Once normalized to unity, it can be expressed as
\begin{eqnarray} \label{eq:ESR}
{\cal R}(\nu,B_{\rm NV})&=&1-{\cal C}\times\Bigg[{\cal G}\left(\frac{\nu-\nu_{R,+}(B_{\rm NV})}{\Delta\nu/2\sqrt{\ln2}}\right) + \\
& &{\cal G}\left(\frac{\nu-\nu_{R,-}(B_{\rm NV})}{\Delta\nu/2\sqrt{\ln2}}\right)\Bigg]
\end{eqnarray}
where ${\cal G}(x)=\mathrm{e}^{-x^2}$ denotes a Gaussian function, $\Delta\nu$ is the ESR linewidth (FWHM) and ${\cal C}$ its contrast. Two measured ESR spectra are shown in Figure~\ref{fig:ESR_vs_B}(c) for a single NV defect attached to the AFM tip, together with data fitting using Equation~(\ref{eq:ESR}). Varying $B_{\rm NV}$ enables us to verify Equation~(\ref{eq:ESRfreq3}) experimentally, as illustrated in Figure~\ref{fig:ESR_vs_B}(d). Note that the vanishing slope at very low magnetic fields -- i.e. when $\gamma_e B /2\pi \ll E$ -- is responsible for the fact that in our dual-iso-B images, the iso-B lines corresponding to zero field are always wider that those corresponding to a non-zero field. 

\subsection{Determining the orientation of the NV defect}

\begin{figure}[h!]
\begin{center}
    \includegraphics[width=0.49\textwidth]{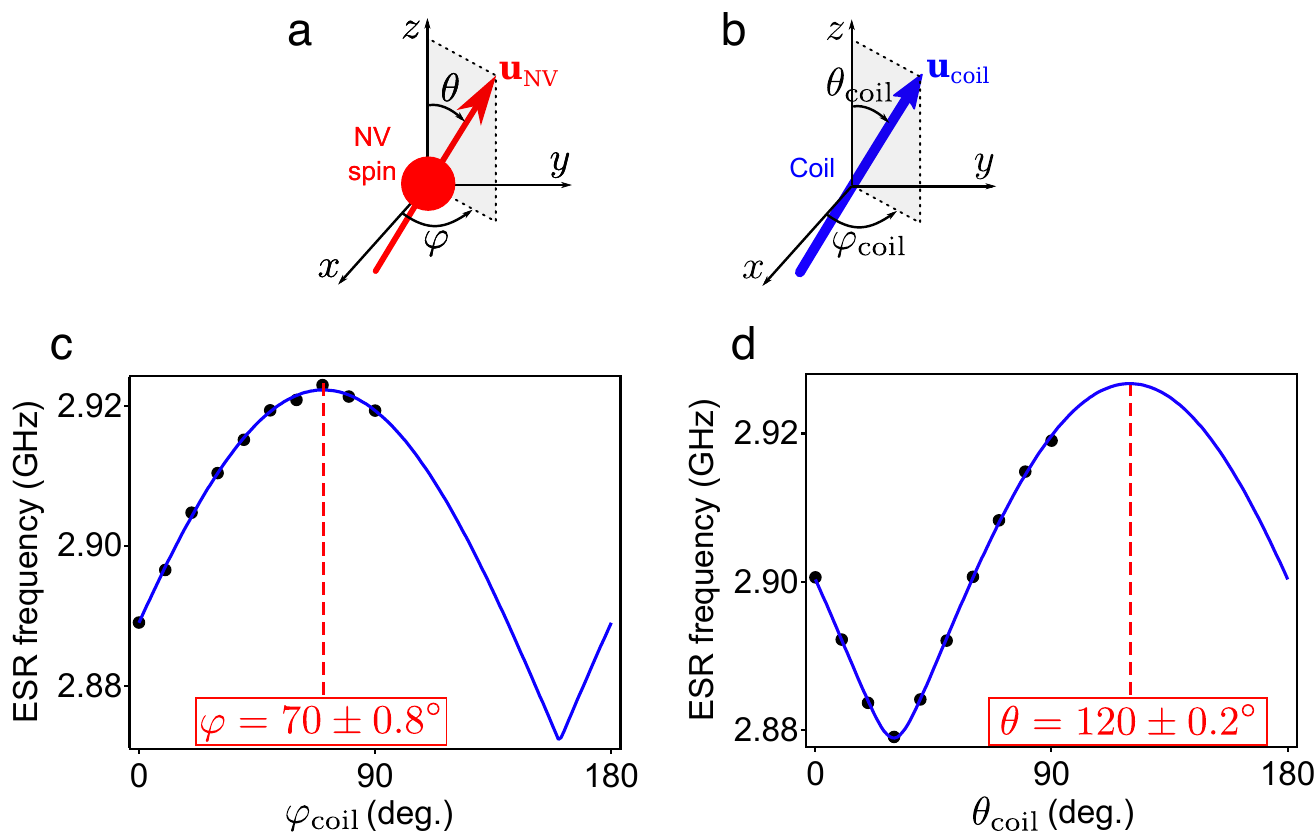}
\end{center}
\caption{(a)-The axis of the NV defect ${\bf u}_{\rm NV}$ is described by the angles $\theta$ and $\varphi$ in the spherical coordinate system of the laboratory reference frame. (b)-An external magnetic field is applied using a three-axis coil system. Its direction, of unit vector ${\bf u}_{\rm coil}$, is described by the angles $\theta_{\rm coil}$ and $\varphi_{\rm coil}$ in the same coordinate system. (c)-Measured ESR frequency $\nu_{R,+}$ as a function of the angle $\varphi_{\rm coil}$ of the applied magnetic field, while fixing $\theta_{\rm coil}=90^\circ$. The solid line is a fit to the data using Equation~(\ref{eq:ESRfreq3}), which yields the angle $\varphi$. (d)-Measured ESR frequency $\nu_{R,+}$ as a function of the angle $\theta_{\rm coil}$ while keeping $\varphi_{\rm coil}=\varphi$. The solid line is a fit to the data using Equation~(\ref{eq:ESRfreq3}), which yields the angle $\theta$. }
\label{fig:NVorientation}
\end{figure}

The orientation ${\bf u}_{\rm NV}$ of the NV defect gives the projection axis of the measured magnetic field in all magnetometry experiments, {\it i.e.} $B_{\rm NV}={\bf B}\cdot {\bf u}_{\rm NV}$. To determine the spherical angles $\theta$ and $\varphi$ that characterize ${\bf u}_{\rm NV}$ [Fig.~\ref{fig:NVorientation}(a)], we apply a calibrated magnetic field ${\bf B}_{\rm coil}=B_{\rm coil} {\bf u}_{\rm coil}$ using a three-axis coil system [Fig.~\ref{fig:NVorientation}(b)] while monitoring one of the two ESR frequencies of the NV defect ({\it e.g.} $\nu_{R,+}$). The procedure we employed is as follows. We first set $B_{\rm coil}=2$ mT and $\theta_{\rm coil}=90^\circ$, and sweep the angle $\varphi_{\rm coil}$ from 0 to 90$^\circ$. This yields the angle $\varphi$ by fitting Equation (\ref{eq:ESRfreq3}) to the data, using the fact that $B_{\rm NV}\propto\cos(\varphi_{\rm coil}-\varphi)$ [Fig. \ref{fig:NVorientation}(c)]. We next set $\varphi_{\rm coil}=\varphi$ and sweep $\theta_{\rm coil}$ from 0 to 90$^\circ$. The angle $\theta$ is deduced by fitting Equation~(\ref{eq:ESRfreq3}) to the data, given that $B_{\rm NV}\propto\cos(\theta_{\rm coil}-\theta)$ [Fig. \ref{fig:NVorientation}(d)]. This procedure enables us to determine the angles $\theta$ and $\varphi$ in the range $[0,\pi]$, which completely defines the direction of ${\bf u}_{\rm NV}$.  \\

\subsection{Magnetization and stray field simulations}

\begin{figure}[h!]
\begin{center}
    \includegraphics[width=0.5\textwidth]{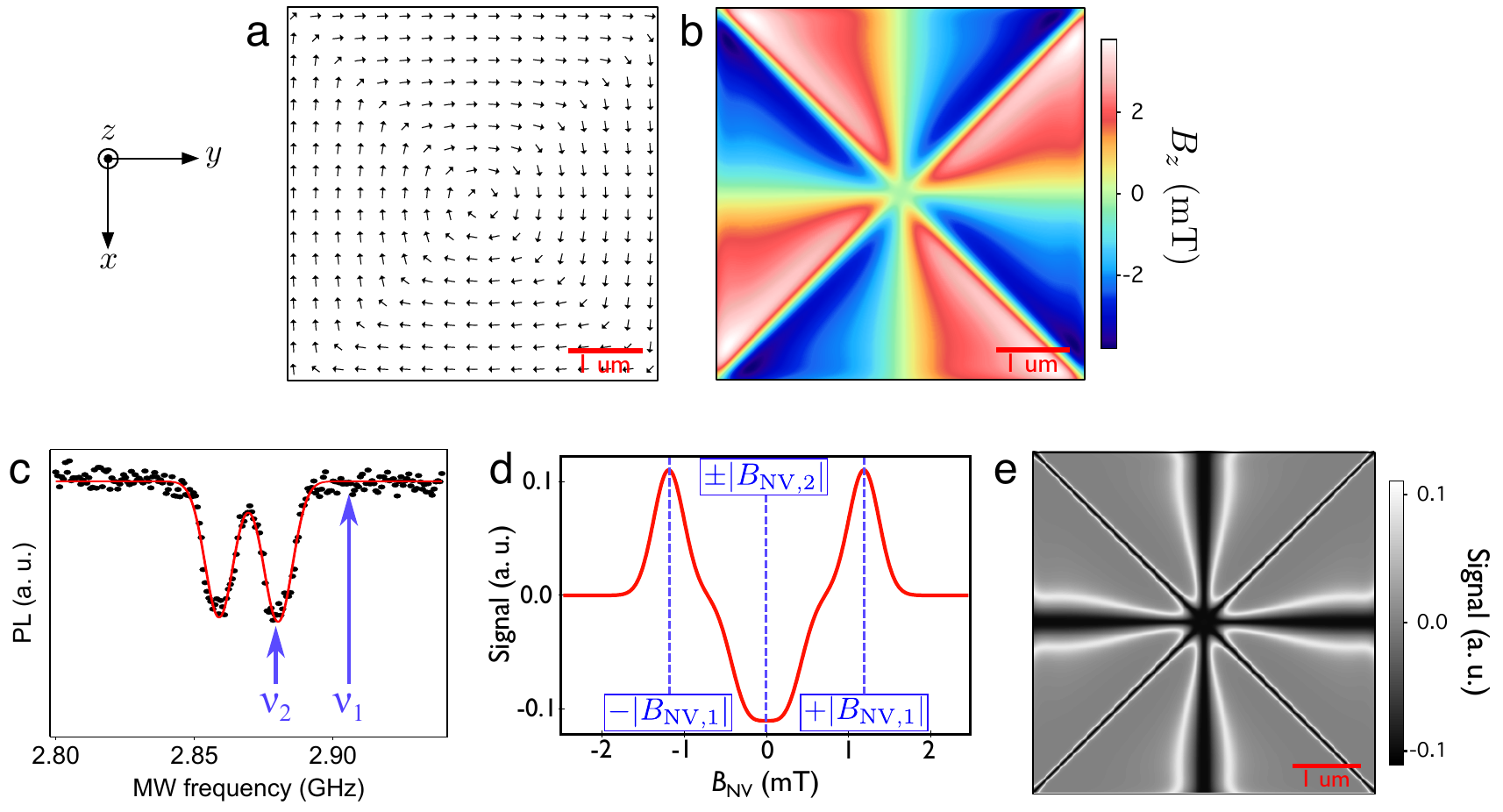}
\end{center}
\caption{(a)-Map of the in-plane magnetization of a 5 $\mu$m Fe$_{20}$Ni$_{80}$ square dot, as computed by OOMMF software. (b)-Calculated distribution of the $B_z$ component of the stray magnetic field at a distance $d=200$ nm above the sample surface. (c)-ESR spectrum measured at $B_{\rm NV}=0$. The microwave frequencies applied for the dual-iso-B images are chosen to be $\nu_1=2905$ MHz and $\nu_2=2880$ MHz, as indicated with blue arrows. (d)-PL difference signal ${\cal S}(\nu_1,\nu_2,B_{\rm NV})$ as a function of $B_\mathrm{NV}$, calculated from the ESR spectrum in (c) and Equation (\ref{eq:dual}). (e)-Dual-iso-B image calculated by applying ${\cal S}(\nu_1,\nu_2,B_{\rm NV})$ to the magnetic field map shown in (b). Here we assumed the NV axis to be parallel to the $z$ axis.}
\label{fig:simu_MB}
\end{figure}

The magnetization distribution inside the square film of Fe$_{20}$Ni$_{80}$ was computed using OOMMF software\cite{oommfbis}. The parameters used for Fe$_{20}$Ni$_{80}$ were: spontaneous magnetization $M_s=8 \, 10^{5}$A/m and an exchange constant $A=10^{-11}$~J/m. A magnetization cell size $5\times 5\times 5 \ {\rm nm}^{3}$ was used, which is below the micromagnetic exchange length in order to describe the vortex core structure precisely. From the calculated equilibrium magnetization [Fig. \ref{fig:simu_MB}(a)], the stray magnetic field is computed by summing the contribution of all magnetization cells, and then projected along the NV defect axis in order to reach a map of $B_{\rm NV}$. As an example, Figure \ref{fig:simu_MB}(b) shows the calculated $z$-axis component of the stray field for a probe-to-sample distance $d=200$~nm.

In the following, we take into account the strain-induced splitting, which was omitted in the main text for clarity purpose. We therefore use Equation~(\ref{eq:ESRfreq3}) for the simulation of the magnetic field maps recorded with the scanning NV magnetometer. The dual-iso-B images shown in the main article are obtained by measuring the difference of NV defect PL intensity for two fixed MW frequencies $\nu_1$ and $\nu_2$. The resulting signal reads
\begin{equation} \label{eq:dual}
{\cal S}(\nu_1,\nu_2,B_{\rm NV})={\cal R}(\nu_2,B_{\rm NV})-{\cal R}(\nu_1,B_{\rm NV}) \ ,
\end{equation}
where ${\cal R}(\nu,B_{\rm NV})$ is given by Equation~(\ref{eq:ESR}). The signal ${\cal S}$ is positive (resp. negative) when $\nu_1$ (resp. $\nu_2$) matches one of the two ESR frequencies $\nu_{R,\pm}$ -- {\it i.e.} when $B_{\rm NV}=B_{\rm NV,1}=\pm 2\pi\sqrt{(\nu_1-D)^2-E^2}/\gamma_e$ (resp. $B_{\rm NV,2}=\pm 2\pi\sqrt{(\nu_2-D)^2-E^2}/\gamma_e$) -- and is null otherwise. 

As an illustration, we consider the same NV defect as in Figures~\ref{fig:ESR_vs_B}(c)\&(d), with an ESR spectrum at zero magnetic field reproduced in Figure~\ref{fig:simu_MB}(c). We choose $\nu_1=2905$ MHz and $\nu_2=2880$ MHz, which corresponds to $B_{\rm NV,1}=1.2$ mT and $B_{\rm NV,2}=0$. Having fully characterized the ESR response, {\it i.e.} its linewidth and contrast, the resulting dual-iso-B signal ${\cal S}(\nu_1,\nu_2,B_{\rm NV})$ can be computed as a function of $B_{\rm NV}$, as shown in Figure~\ref{fig:simu_MB}(d). The simulated magnetic field maps are then obtained by applying the ${\cal S}(\nu_1,\nu_2,B_{\rm NV})$ response to the calculated stray field once projected along the NV defect axis. An example of a calculated dual-iso-B image is shown in Fig.~\ref{fig:simu_MB}(e), where we assumed $d=200$ nm and the NV axis to be along the $z$ axis, {\it i.e.} $Z=z$.

\subsection{Distance dependence}

\begin{figure}[h!]
\begin{center}
    \includegraphics[width=0.48\textwidth]{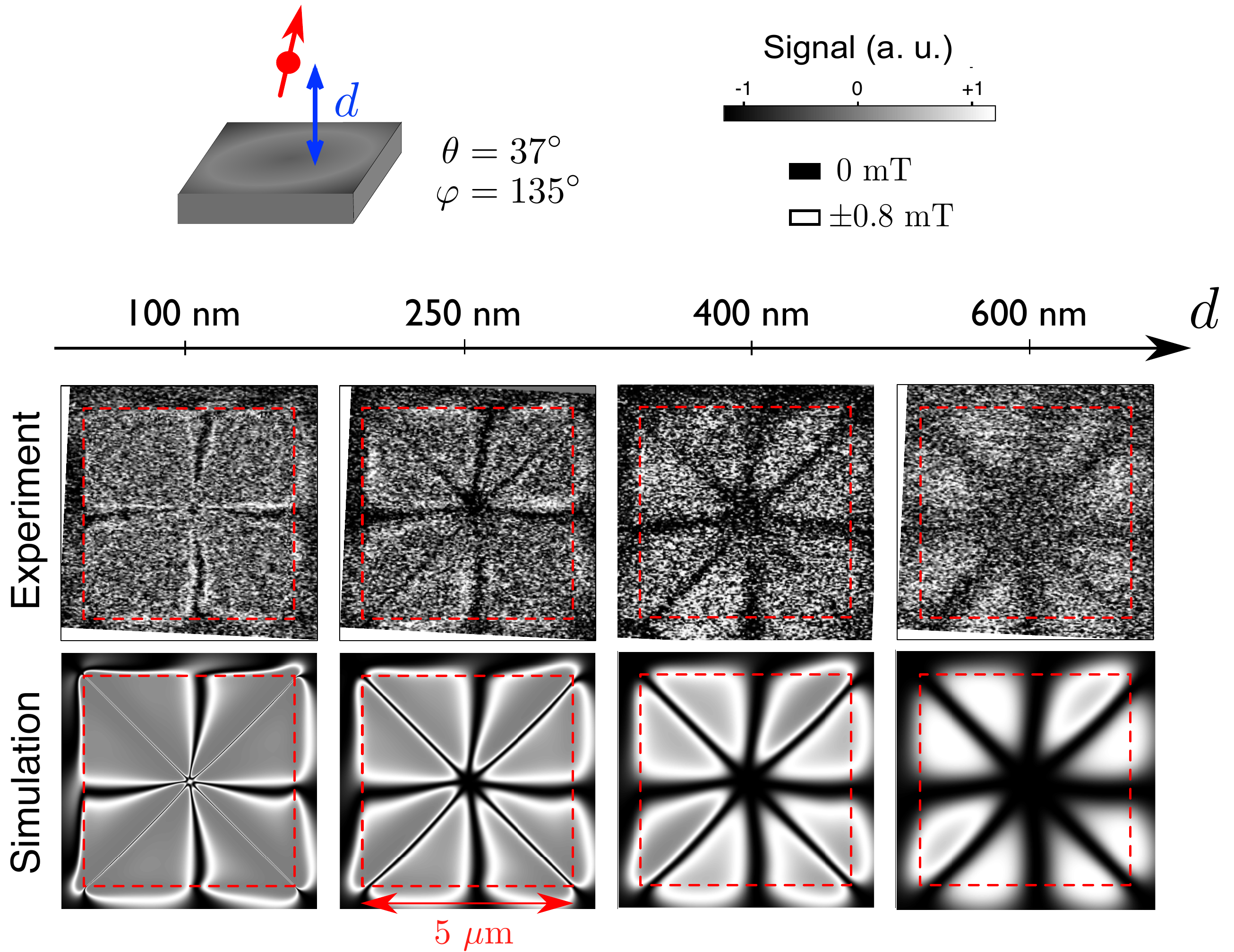}
\end{center}
\caption{Experimental dual-iso-B images (top panels) obtained for a NV defect oriented mainly along the $z$ axis ($\theta=37^\circ$, $\varphi=135^\circ$) at different probe-sample distances $d$, together with the simulations (bottom panels). The red square depicts the position of the 5 $\mu$m square dot.}
\label{fig:BvsZ}
\end{figure}

Magnetic field  images recorded above a 5 $\mu$m Fe$_{20}$Ni$_{80}$ square dot at several probe-to-sample distances, varying between $d=100$ and $d=600$~nm, are shown in Figure~\ref{fig:BvsZ}. For these experiments, the probe-to-sample distance is controlled by using the AFM in a dual-pass mode with a constant lift height during the second pass. The same NV defect probe as in Figure 3 of the main article is used, {\it i.e.} mainly oriented along the $z$-axis. A good agreement with the simulations is found, which illustrates the three-dimensional stray field mapping capabilities of scanning NV magnetometry. The slight asymmetries observed in both the experimental and simulated images result from the imperfect alignment of the NV defect axis with respect to the $z$-axis. 

\subsection{Magnetic-field-induced fluorescence quenching}

\begin{figure}[h!]
\begin{center}
    \includegraphics[width=0.47\textwidth]{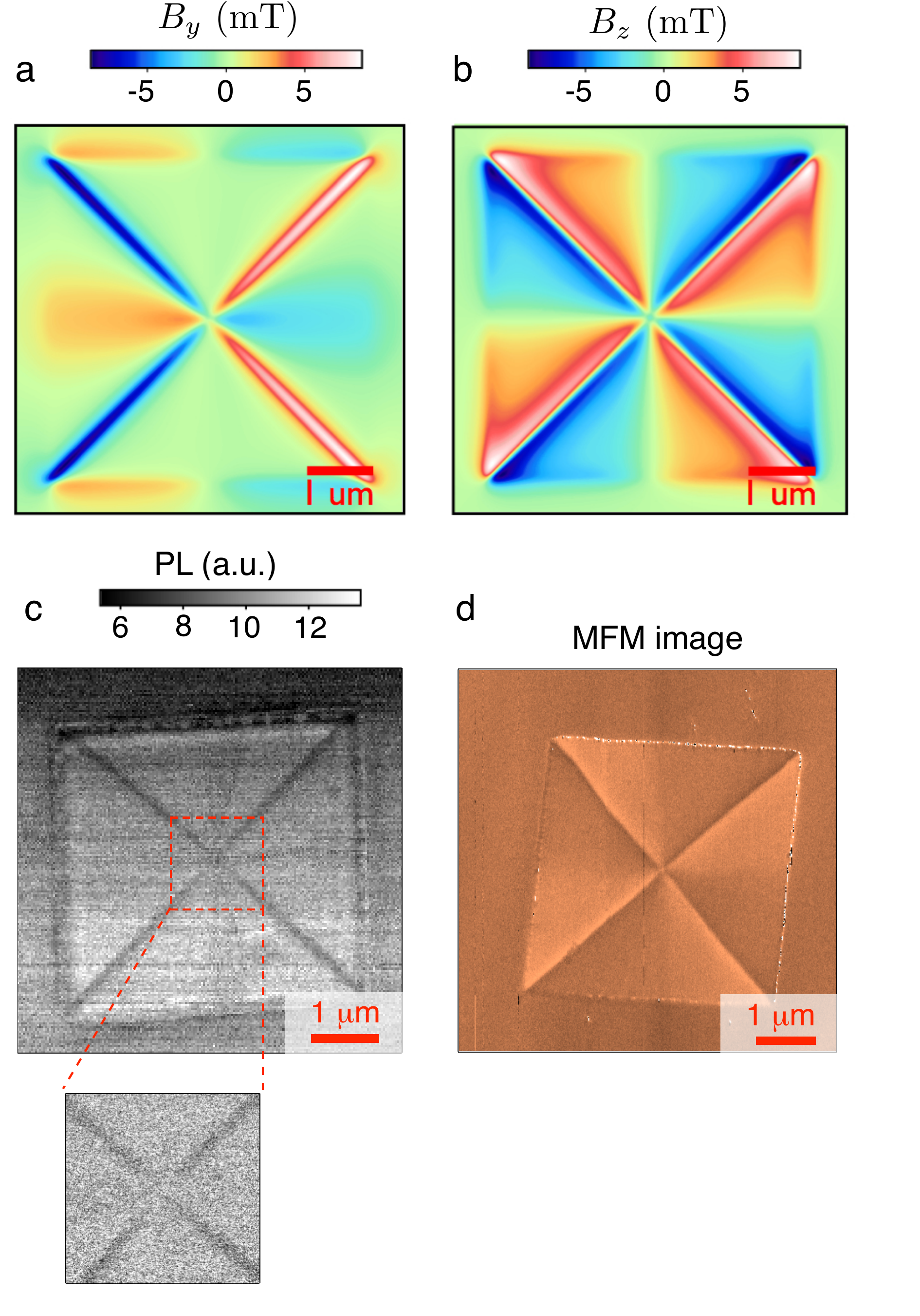}
\end{center}
\caption{(a),(b)-Calculated $B_y$ and $B_z$ components of the stray field distribution at a distance $d=100$~nm. (c)-PL quenching image obtained without applying any MW field and with the AFM operating in tapping mode. Zooming into the center shows that the PL quenching vanishes in the vicinity of the center of the structure, where the magnetic field amplitude is expected to decrease. (d)-MFM image of a similar square of Fe$_{20}$Ni$_{80}$.}
\label{fig:quenching}
\end{figure}

Although the simulations fairly agree with experimental data, the sharp structures lying along the square diagonals are not observed in the experimental image recorded at the smallest probe-to-sample distance $d=100$~nm [Fig.~\ref{fig:BvsZ}]. As explained in the main text, a strong stray field is generated by N\'eel domain walls in these regions. Indeed, the calculated magnetic field maps at $d=100$~nm indicate that the stray field components along the $y$- and $z$-axis reach nearly $\sim 10$~mT at the square diagonals, as illustrated in Figures~\ref{fig:quenching}(a)\&(b). In this condition, the approximation $\mathcal{H}_\perp\ll  \mathcal{H}_\parallel$ is not fulfilled [see Eq.~(\ref{eq:hamiltonian2})] and the quantization axis is determined by the local magnetic field rather than by the NV defect axis. As a result, the spin projection along the NV axis $m_s$ is no longer a good quantum number and the eigenstates of the spin Hamiltonian are given by superpositions of the $m_s=0$ and $m_s=\pm1$ spin sublevels~\cite{Epstein2005}. In this regime, the contrast of optically detected ESR vanishes~\cite{Tetienne2012bis}. Magnetic field imaging through optically-detected ESR is therefore inefficient in the regime of `high' off-axis magnetic field. This explains why the zero-field lines are not visible at the square's diagonals for  $d=100$~nm, where a strong off-axis magnetic field is experienced by the NV defect.\\

It was also shown that the decreased ESR contrast is accompanied by a reduction of the NV defect PL intensity when the magnitude of $B_\perp$ increases~\cite{Epstein2005,Lai2009}. Such a magnetic-field-dependent PL can be exploited as an alternative method to map high off-axis magnetic field regions~\cite{Rondin2012bis,Tetienne2012bis}. This is illustrated in Figure~\ref{fig:quenching}(c) where dark lines are observed along the square's diagonals in the PL image for a probe-to-sample distance $d=100$~nm. In this experiment, we record the NV defect PL intensity without applying any MW field. The dark lines indicate a PL quenching induced by the off-axis component of the magnetic field generated by N\'eel domain walls. Near the center of the magnetic structure, the magnetic field amplitude decreases [see Fig.~\ref{fig:quenching}(c)], which enabled us to image the vortex core using the ESR response of the NV defect. \\

We note that the resulting PL quenching image exhibits obvious similarities with the one recorded with a magnetic force microscope (MFM), which is sensitive to the field gradient [Fig.~\ref{fig:quenching}(d)]. \\

\label{sec:quenching}

\subsection{Vortex polarity and chirality}

\begin{figure*}[t]
\begin{center}
    \includegraphics[width=.6\textwidth]{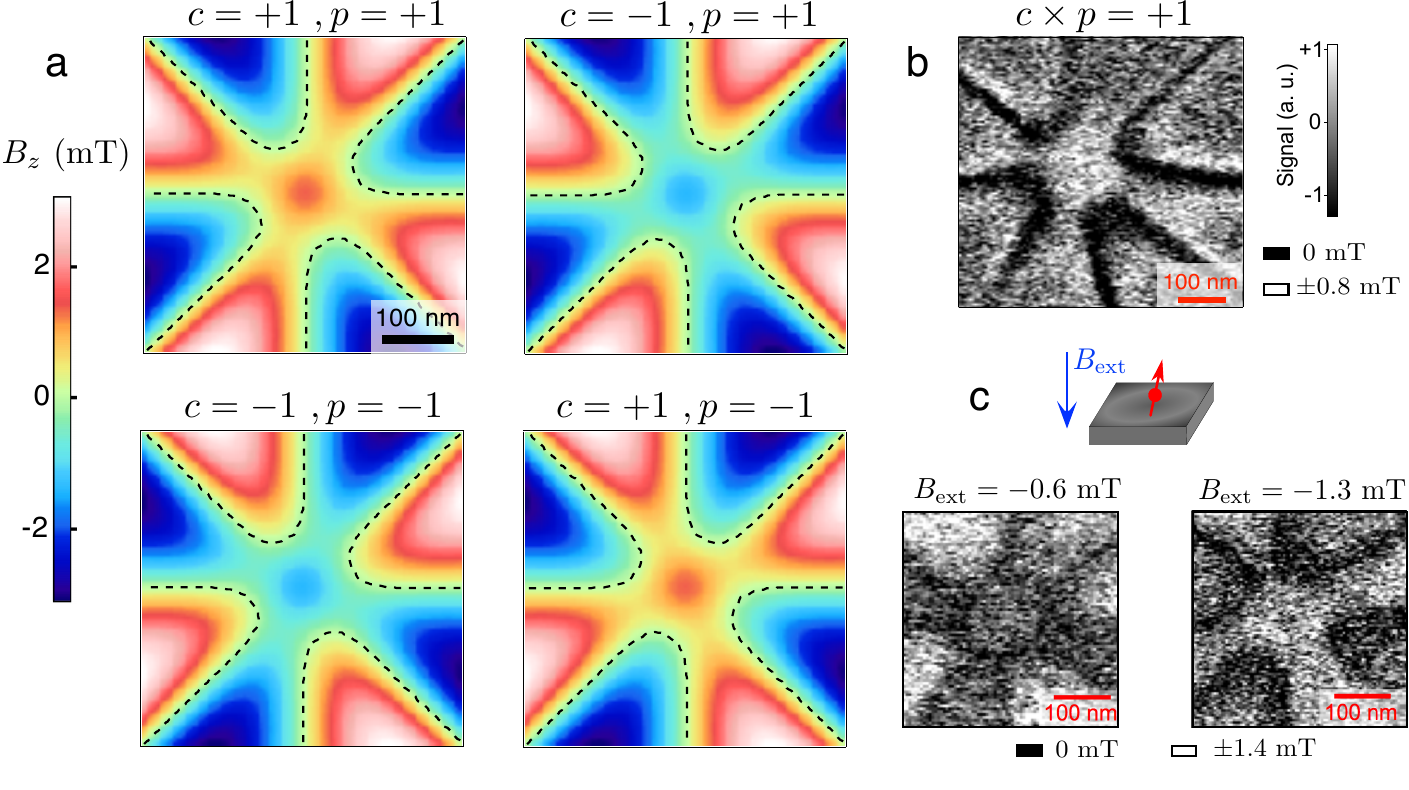}
\end{center}
\caption{(a)-Calculated $B_z$ component of the stray field produced by the center of the magnetic structure at $d=100$~nm, for the four possible combinations of chirality $c$ and polarity $p$ of the vortex state. The black dotted lines indicate the zero-field contours, separating positive (upward) and negative (downward) field regions. (b)-Magnetic field map of the stray field emanating from the vortex core magnetization, as shown in Figure 3 of the main text. (c)-Dual-iso-B image of the square center while applying a bias field with a projection along the NV axis $B_\mathrm{ext}=-0.6$ mT and $B_\mathrm{ext}=-1.3$ mT. }
\label{fig:chir}
\end{figure*}

The product of the vortex chirality $c$ and the vortex core polarity $p$ can be determined by simply examining the zero-field contours around the vortex core in the experimental dual-iso-B images. Indeed, zero-field lines separate the regions of positive and negative magnetic field projection. It is then possible to determine which ``lobes'' have the same sign as the vortex core field. Figure~\ref{fig:chir}(a) shows the magnetic field map calculated at $d=100$~nm from the sample surface and projected along the $z$-axis, for all possible combinations of $c$ and $p$. As expected, the pairing of the zero-field contours around the vortex core (black dotted lines) are fixed by the product $c\times p$. Comparing the experimental results to simulations directly enables to determine that $c\times p=+1$ for the vortex state studied in Figure 3 of the main text [reproduced in Fig.~\ref{fig:chir}(b)]. \\

By applying an external magnetic field $B_{\rm ext}$ that cancels out the net field above the vortex core, the polarity of the vortex core can be determined as shown in the main paper. In Figures~\ref{fig:chir}(c), we rather show dual-iso-B images obtained for a bias field $B_{\rm ext}$ antiparallel to the vortex core's field $B_{\rm core}$ such that $|B_{\rm core}|-|B_{\rm ext}|$ is slightly positive or slightly negative. In the latter case, the net field above the vortex core is now of opposite sign compared to the zero-field case, which reverses the pairing of the zero-field lines in the image. \\

\section{Micromagnetic simulations for domain wall imaging}

As indicated in the conclusion of the main manuscript, the direct quantitative measurement of weak magnetic fields, without perturbation and with a nanoscale resolution, may allow clearing some important issues in nanomagnetism. As an example, we show in this final section how scanning-NV magnetometry might allow to clearly identify the nature of magnetic domain walls (DW) in ultrathin ferromagnetic films with perpendicular anisotropy.
\begin{figure}[b]
\begin{center}
\includegraphics[width=0.45\textwidth]{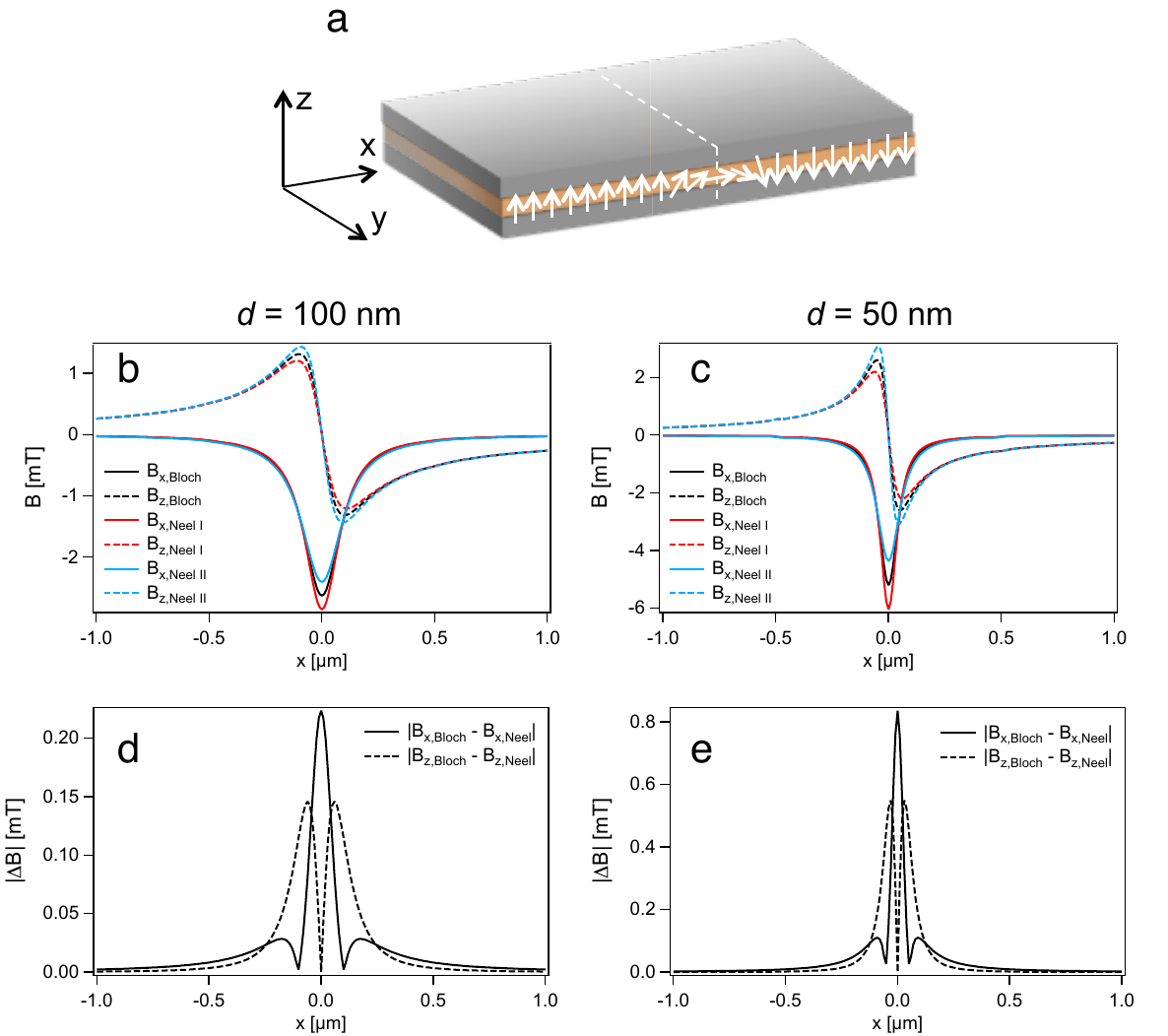}
\end{center}
\caption{{\bf Simulated stray field of a domain wall (DW).} (a)-Scheme of a domain wall in an ultrathin ferromagnetic film with perpendicular anisotropy in contact with non magnetic layers. (b),(c)-Stray field generated by a DW in a Bloch configuration or in a N\'eel configuration at (b) $d=100$~nm and (c) $d=50$~nm from the sample. The notations N\'eel I and N\'eel II indicate the different N\'eel wall chirality. (d),(e)-Absolute difference of the stray field between a Bloch and a N\'eel domain wall, at (d) $d=100$~nm and (e) $d=50$~nm from the sample. The sign of the difference is given by the N\'eel wall chirality. }
\label{Fig_linecutDW}
\end{figure}

The magnetism of ultrathin films has attracted a renewed interest since the experimental discovery of the relevance of Dzyaloshinskii-Moriya interactions (DMI) in ultrathin magnetic films in contact with non magnetic layers~\cite{Bode2007}. In particular, it has been shown that the DMI affects the nature of the lowest-energy magnetic domain wall structures in such films \cite{Thiaville2012bis}, which in turn controls their current-induced motion~\cite{Khvalkovskiy2013}. In the following, we present micromagnetic simulations of domain walls (DWs) in a thin film with perpendicular anisotropy, the typical sample being Pt/Co(0.6nm)/AlO$_x$~\cite{Miron2011}. 

As shown in Figure~\ref{Fig_linecutDW}(a), we consider a film parallel to the $x$-$y$ plane, with a DW centered at $x=0$ that extends to infinity along the $y$ direction. The  magnetization distribution in the case of a Bloch- and N\'eel-type DW was obtained using OOMMF software, from which the magnetic stray field was computed [Figure~\ref{Fig_linecutDW}(b)\&(c)] with typical probe-to-sample distances $d=100$~nm and $d=50$~nm. Figure~\ref{Fig_linecutDW}(d)\&(e) shows the absolute difference in the magnetic stray field produced by the two types of DWs at the same probe-to-sample distances. Obviously the calculated stray fields, reaching a few milliteslas right above the DW, are in the operation range of the NV magnetometer, and the $\pm 10$\% difference between the two types of DW is large enough to allow the NV magnetometer to experimentally determine the nature of the DW.



\begin{thebibliography}{30}
\bibitem{Note}
Higher-energy magnetization states might have appeared during the sample growth. We verified that applying a strong rotating magnetic field restored all ferromagnetic squares in their minimum-energy single-vortex state.
\end{thebibliography}
\end{document}